\documentclass{aa}
\input psfig.sty
\input epsf.sty
\begin{document}

   \title{Self-similarity of clusters of galaxies and the $L_X-T$ relation}

   \author{D.M. Neumann, M. Arnaud       }

   \offprints{D.M. Neumann}

   \institute{CEA/Saclay DSM/DAPNIA/ Service d'Astrophysique,
              L'Orme des Merisiers, B\^at. 709, F-91191 Gif-sur-Yvette,
              France\\
              email: ddon@cea.fr , marnaud@discovery.saclay.cea.fr
}
    \date{}

\def \kT  {{\rm k}T}
\def \keV  {\rm keV}
\def \etal      {et al.\ }
\def \LxT {\hbox{$L_{\rm X}$--$T$} }
\def \Mg {M_{\rm gas}}
\def \MgT {\hbox{$M_{\rm gas}$--$T$} }
\def \betamodel {\hbox{$\beta$--model} }

\abstract{ In this paper based on ROSAT/PSPC data we investigate the emission 
measure profiles  of a sample of hot clusters of galaxies ($kT>3.5$~keV) in 
order to explain the differences between observed and theoretically predicted 
\LxT relation. Looking at the form of the emission measure profiles as well as 
their normalizations we find clear indication that indeed  the profiles 
have similar shapes once scaled to the virial radius, however, the 
normalization of the profiles shows a strong temperature dependence. We 
introduce a \MgT relation with the dependence  $\Mg \propto 
T^{1.94}$. This relationship explains the observed \LxT relation
and reduces the scatter in the scaled
profiles by a factor of 2 when compared to the classical scaling.  We
interpret this finding as strong indication that the \MgT relation in
clusters deviates from classical scaling.}


\maketitle
      \keywords{Galaxies: clusters: general --
                Cosmology: observations, dark matter --
                X-rays: general
               }

%

\section{Introduction}

The simplest, purely gravitational, models of structure formation predict that
galaxy clusters are self-similar in shape and that scaling laws
relate each physical property to the cluster total mass $M$ and redshift $z$ 
(Kaiser \cite{kaiser1}; Navarro et al. \cite{navarro}; Teyssier et al. 
\cite{teyssier}; Eke et al.\cite{eke}; Bryan \& Norman \cite{bryan}).  
Self similarity applies to both
 the dark matter component and the hot intra-cluster medium (ICM).  The virial 
theorem then yields the well known scaling relations:
 \begin{eqnarray}
M_{\rm \delta}&\propto& T^{3/2}(1+z)^{-3/2}\\
R_{\rm \delta}&\propto& T^{1/2}(1+z)^{-3/2}.
\end{eqnarray}
where $M_{\rm \delta}$ is the total mass in the sphere of radius $R_{\rm
\delta}$ corresponding to the overall over density $\delta$ and $T$ is
the cluster temperature.  Assuming a constant gas mass fraction, the
total gas mass then scales as $\Mg \propto T^{3/2}$ and the
X--ray luminosity as $T^{2}$.

The comparison between observed clusters and theoretical predictions gives us
insight into the physics that governs their formation and
evolution.  Vikhlinin et al. \cite{vikhlinin} recently found $R \propto
T^{0.57\pm0.04}$, which is close to the predicted scaling law.  Both
the gas density and temperature profiles of hot clusters ($T > 4 $keV)
do show regularity (Markevitch et al. \cite{markevitch1}; Neumann \& Arnaud 
\cite{neumann},
hereafter paper I; Vikhlinin et al. \cite{vikhlinin}).  The shapes of
various clusters, once the radius is scaled to the virial radius, look
remarkably similar outside the cooling flow region, supporting the
existence of an universal underlying dark matter profile.

However, clusters also deviate from the predicted scaling laws.  The
most remarkable deviation is the observed \LxT relation.  Different
authors, studying different cluster samples, found $L_X \propto
T^{2.9}$ (Arnaud \& Evrard \cite{arnaud}; Markevitch \cite{markevitch2}; 
Allen \& Fabian \cite{allen}) with a low dispersion, ruling out the 
theoretical relation $L_X \propto T^2$.  This difference between the slopes of 
the observed and theoretical relations can be explained if i) the cluster 
shape depends
on the ICM temperature and/or ii) the gas mass temperature relation
deviates from predictions (Arnaud \& Evrard \cite{arnaud}; Ponman \etal 
\cite{ponman}).
Non gravitational effects, like pre-heating by early galactic winds,
could affect the gas distribution shape (e.g Loewenstein \cite{loewenstein}; 
Tozzi \& Norman \cite{tozzi}) and have been proposed to explain the steepening
of the \LxT relation (Kaiser \cite{kaiser2}; Evrard \& Henry \cite{evrard1}). 
 However the
observed structural similarity of hot clusters suggest that pre-heating
plays an important role only for cool ($T \leq 3.5~{\rm keV}$) clusters
(see also Ponman \etal \cite{ponman}, Lloyd-Davis \etal \cite{lloyd}).

In order to investigate further the observed \LxT relation, we examine
in this paper the emission measure profiles of a sample of nearby hot
clusters, for which the effect of pre-heating is a priori minimal.  The
shape and normalization of the emission measure profiles are sensitive
to both the cluster internal structure and the total gas mass.
Comparing the profiles of clusters at different temperatures can
provide more detailed information about the cause of the discrepancy
between the predicted and observed \LxT relations.  In this paper we adopt
$H_0 = 50 {\rm km/sec/Mpc}$ and $\Omega_{\rm m}=1$.


\section{The data}
\label{sec:sample}

The emission measure profiles are from our previous work (Paper I),
where the cluster sample and imagery data processing are described in
detail.  The cluster sample comprises Abell clusters (Abell, Corwin \&
Olowin \cite{abell}) in the redshift range $0.04<z<0.06$, which were observed
in pointing mode with the ROSAT PSPC and provide good statistics.
Here we only consider the clusters in the subsample for which accurate
temperature measurements exist from the literature (with $\kT >
3.7\keV$).  The emission measure profiles are deduced from the
observed surface brightness profiles via Eq.3 in paper I.

For the \LxT relation, we adopt here and in the following
the relation derived by Arnaud \& Evrard (\cite{arnaud}):
\begin{equation}
L_X \propto T^{2.88}
\label{equ:lxt288}
\end{equation}
which is in good agreement with other works (Markevitch \cite{markevitch2}; 
Allen \& Fabian \cite{allen}).

\section{Combining self-similarity and the observed $L_X-T$-relation}

\subsection{Dependence of X-ray luminosity and
emission measure profile on gas mass and ICM structure}

Let us consider a cluster of extent $R$ and temperature $T$. Its X-ray
luminosity depends on its total gas mass (within radius $R$) and
its internal structure as:
\begin{equation}
L_X \propto \frac{\Mg^2}{R^3} \Lambda(T) Q(T)
\label{equ:lxt}
\end{equation}    
(see also Arnaud \& Evrard \cite{arnaud}) where $\Lambda(T)$ is the cooling 
function, which we will 
assume to be $T^{1/2}$. $Q(T)$ is a form factor
which only depends on the shape of the gas density distribution
($\rho_{gas}(r)$).  $Q(T)$ is equal to $\langle \rho^2_{gas}\rangle /
\langle \rho_{gas}\rangle^2$, where the brackets denote the average
over the whole cluster (see also Arnaud \& Evrard \cite{arnaud}).  If clusters
are self-similar in shape, $Q(T)$ should be constant.

The emission measure along the line-of-sight at radius $r$ is defined
as:
\begin{eqnarray}
EM(r) \propto \int_r^{R} \frac{n_g^2(x)xdx}{\sqrt{x^2-r^2}}
\propto
\frac{\Mg^2}{R^5}F(r/R,T)
\label{equ:em}
\end{eqnarray}
$F(r/R,T)$ is again a dimensionless form function, which should be independent
of temperature if clusters are self-similar. According to the
theoretical scaling laws, EM should scale as $T^{1/2}$.

\subsection{Does the ICM shape depend on temperature?}

\begin{figure}
\psfig{figure=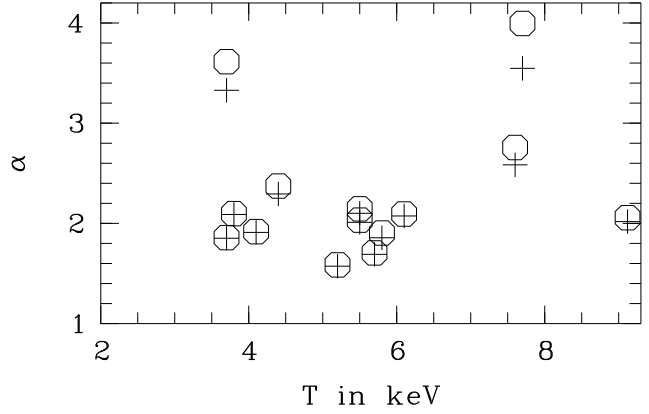,width=8.5cm}
\caption{The $\alpha$ parameter (logarithmic density gradient) versus
temperature.  The crosses show $\alpha$ at a scaled radius of r/R=0.3,
the circles at r/R=0.6.  }
\label{fig:alpha}
\end{figure}

In Paper I, we already quantified the structural variation in the
cluster sample.  We derived the logarithmic ICM density gradient
$\alpha$ by fitting a $\beta$--model to the data,
\begin{equation}
\alpha(r/R) =$ $ -
\frac{d \log n_g(r/R)}{d \log (r/R)}= 3 \frac{\beta}{1+(r_c/r)^2}.
\end{equation}
and showed that the dispersion of $\alpha$ is less than $20\%$ at any
scaled radius.   The $\alpha$
values, for two different scaled radii, are plotted versus cluster
temperature in Fig.\ref{fig:alpha}.  There is no apparent correlation
between $\alpha$ and $T$ , i.e no systematic variation of shape with
temperature.

Fig.\ref{fig:class} shows the scaled emission measure profiles of the
clusters, $\widetilde{EM}(r/R) \propto EM(r)/T^{1/2}$ (upper panel),
and their corresponding relative dispersion (lower panel).  The
scaling radius was fixed to $R = r_{200}$ with the normalization from
Evrard et al. (\cite{evrard2}).  Due to detection limits we only
consider  radii $r/R<0.5$, our sample becoming incomplete at
larger radii.  The dispersion of the profiles is significant ($40\%$
standard deviation) but stays constant with radius.  This again
indicates that the profiles are essentially parallel (self similarity
of form) but points towards a possible problem with the normalization
scaling factor.  These features are also evident to the eye in the
upper panel.

We will thus assume in the following a common shape for all the
clusters in the sample (i.e. $Q(T)$ and $F(r/R,T)$ are constant) and
study possible deviation from the theoretical \MgT and
$R$--$T$ relations.

\subsection{Temperature dependence of the gas mass}

\begin{figure}[t]
\psfig{figure=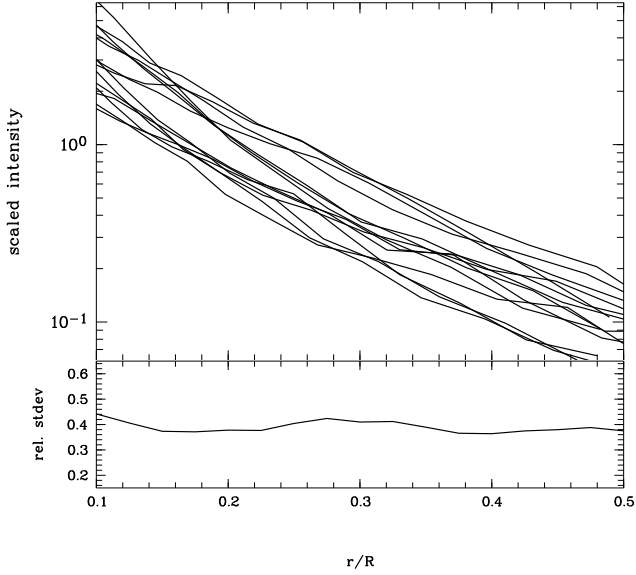,width=8.5cm}
\caption{{\bf Upper panel:}
Scaled emission measure profiles using the theoretical scaling laws
(arbitrary units). For A3562 we used $kT=5.2$keV a new measurement obtained by 
Ettori et al. \cite{ettori}. {\bf Lower panel:} The corresponding standard 
deviation divided by the mean value as a function of scaled radius. }
\label{fig:class}
\end{figure}

We now
consider:
\begin{equation}
R \propto T^x ~~~~~{\rm and}~~~~~
M_{\rm gas} \propto T^{3/2+y}
\label{equ:rvfg}
\end{equation}
From Eq.~\ref{equ:lxt288}, Eq.~\ref{equ:lxt} and Eq.~\ref{equ:rvfg},
the x and y slopes must be related by
$3x - 2y = 0.62$
to account for the observed slope of the $L_X-T$ relation.

\begin{figure}
\psfig{figure=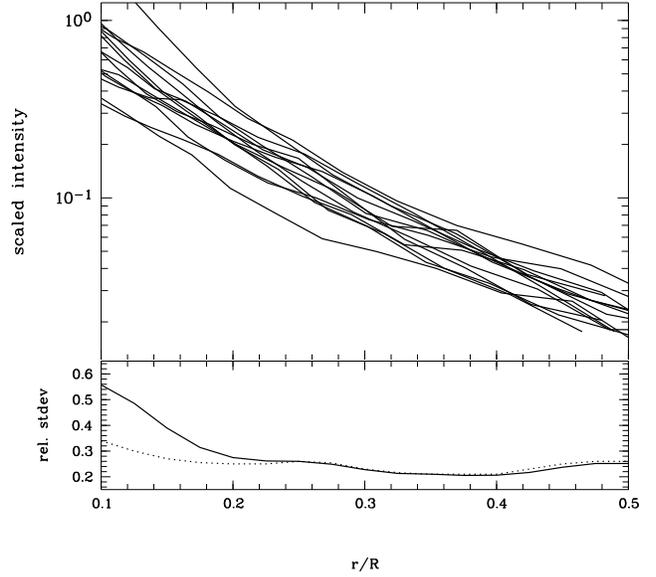,width=8.5cm}
\caption{ {\bf Upper panel:}
Scaled emission measure profiles (arbitrary units) assuming $\Mg
\propto T^{1.94}$ ($x=0.5$, $y=0.44$). For A3562 we used the
$kT$ measurement by Ettori et al. \cite{ettori}. The dynamic range is the same
as in
Fig.\ref{fig:class}.
{\bf Lower panel:} The full line shows the corresponding standard deviation 
divided by the mean value for all clusters. The dotted line shows the relative
dispersion without  A780, which hosts a strong cooling flow and which 
increases the dispersion considerably at small radii.}
\label{fig:T088}
\end{figure}

If we assume that $x=0.5$ , corresponding to the classical $R$--$T$
relation\footnote{Another formal solution is $y=0$, which implies $x = 0.21$. 
This
is in contradiction with the $R-T$ relation found by Vikhlinin \etal 
(\cite{vikhlinin}). Therefore we discard this possibility in the following.}, 
we derive
$y=0.44$ or $ M_{\rm gas} \propto T^{1.94}$.  From Eq.~\ref{equ:em},
the emission measure scales in this case as $EM \propto T^{1.38}$,
instead of $EM \propto T^{1/2}$.  The scaled emission measure
profiles, using this new scaling, are shown in Fig.\ref{fig:T088}.
Note that these rescaled profiles are simply the scaled profiles,
$\widetilde{EM}$, displayed Fig.\ref{fig:class}, divided by
$T^{0.88}$.  Remarkably, this re-normalization does reduce the
dispersion of the profiles, by a factor of 2, when compared to the
classical scaling.

\begin{figure}[h]
\hspace*{0.5cm}
\psfig{figure=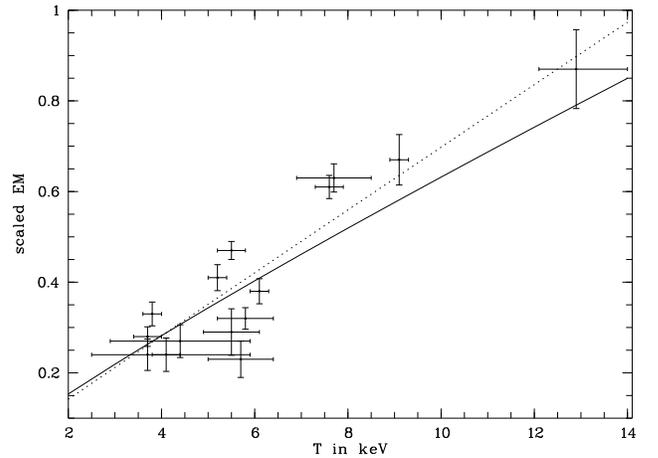,width=8.5cm}
\caption{ The scaled emission
measure of the clusters (classical scaling law) versus temperature at
$r/R=0.3$ (about 1~Mpc).  The error bars for the temperature are
90\% c.l. and for the emission measure 1$\sigma$.  The full line
represents $EM \propto T^{0.88}$ and the dashed line $EM \propto
T^{0.99}$ (see also text).}
\label{fig:T-em}
\end{figure}

To study more directly the deviation from the theoretical temperature
scaling, we display $\widetilde{EM}$, estimated at $r/R=0.3$, versus T
in Fig.\ref{fig:T-em}.  To extend the upper limit of the sample
temperature range, we also considered A2163 ($\kT\sim 13 \keV$), one
of the hottest clusters observed so far (Elbaz \etal \cite{elbaz}).  A clear
correlation between $\widetilde{EM}$ and T is observed.  The
relationship required to explain the observed $L_{\rm X}-T$ relation
($\widetilde{EM}\propto T^{0.88}$, full line) matches the data points.
The dashed line in Fig.~\ref{fig:T-em} represents a crude best fit to
the data (Press \etal \cite{press})
ignoring error bars.  This fit gives $\widetilde{EM}\propto T^{0.99}$,
reasonably close to the required relationship.  The error bars on
$\widetilde{EM}$ and $T$ are highly correlated in a non trivial way
and we did not try to estimate error bars on the correlation.  At
fixed scaled radius, $\widetilde{EM}$ is derived from the $EM$
value, measured at a physical radius which depends on $T$ via the
$R$--$T$ relation.  This can be an important source of error, not
included here, since $EM$ varies rapidly with radius.  The relatively
large dispersion at low temperatures (Fig.\ref{fig:T-em}) is possibly
due to the larger number of clusters in this domain and our
underestimate of the true errors.

\section{Discussion}

Recently Mohr et al. (\cite{mohr2}) estimated the gas mass
within $r_{500}$ for a sample of 45 galaxy clusters.  Their derived
\MgT relation, $\Mg \propto T^{1.98\pm0.18}$, is steeper than the
theoretical expectation and in excellent agreement with our results.
Our results are also consistent with the relation, $\Mg \propto
T^{1.71\pm0.13}$, derived by Vikhlinin et al. (\cite{vikhlinin}) after
fitting a power law to the X-ray emission of the outer parts of
clusters.  Although this general agreement was not unexpected, since all
these studies are based on ROSAT surface brightness data, it is not
entirely trivial.  Our evidence of a steeper than predicted  \MgT relation
relies on structural and scaling properties of the emission
measure profiles, guided by the independently measured \LxT relation,
while the analysis in previous works is more global, based on the
integrated density profiles.

Mohr \& Evrard (\cite{mohr1}) found a tight correlation between the X-ray
isophotal radius and the temperature with $R_{\rm I} \propto
T^{0.93\pm0.11}$.  This relation is steeper than the theoretical
expectation, $R_{\rm I} \propto T^{2/3}$, for self similar clusters
with $\beta=2/3$ (Mohr \etal \cite{mohr2}).  They interpret their finding as
evidence for structural variation with temperature.  However, our study 
indicates that this steepening (for clusters with $kT>4$~keV)
originates from systematic deviation of the profile {\it
normalization} from the theoretical scaling with T. For a \betamodel, the 
X-ray surface brightness profiles
at large radii varies as $I(r) \propto EM_{0} (r/R_{\rm
c})^{1-6\beta}$, where $R_{\rm c}$ is the core radius and $EM_{0}$ the
central emission measure.  For self similar clusters $R_{\rm c}
\propto R_{\delta} \propto T^{1/2}$ and $R_{\rm I}$ scales as
$EM_{0}^{1/3} T^{1/2}$ (for $\beta=2/3$).  If $EM_{0}$ scales as
$T^{1.38}$, instead of $T^{1/2}$, as derived here, we expect that
$R_{\rm I}$ scales as $T^{0.96}$, as found by Mohr \& Evrard (\cite{mohr1}).

The steepening of the \MgT relation has to be explained physically.
It can arise from i) a variation of the gas mass fraction $f_{\rm
gas}$ with temperature and/or ii) a deviation of the total mass versus
T relation from the theoretical scaling law.

If the modeling of dark
matter clustering is correct and the ICM is in hydrostatic
equilibrium, the latter possibility would imply that the X-ray mean
temperature is a biased indicator of the virial temperature.  For
instance relativistic particles creating radio halos could contribute
to the ICM pressure.  If the strengths or quantity of these particles
is a function of temperature as indicated by Colafrancesco 
(\cite{colafrancesco}) it
might explain the deviation from the $M-T$ relation.  Departure from
isothermality could also bias the temperature $T_{X}$ measured from
single temperature fits from global cluster X--ray spectra.  
Numerical simulations (Mathiesen \& Evrard \cite{mathiesen}; Bialek \etal 
\cite{bialek}) suggest a slight steepening of
the $M-T_{X}$ relation with $M\propto T_{\rm X}^{1.6-1.7}$.  New X-ray
observatories such as XMM or Chandra will be able to investigate this
possibility in detail.

The effect of pre-heating on the \MgT relation was studied recently by
Bialek \etal (\cite{bialek}). Numerical simulations with initial entropy level
consistent with the observations of Lloyd-Davis \etal (\cite{lloyd}), predict
a steepening of the \MgT relation, typically $\Mg \propto
T^{1.90\pm0.07}$, which is coherent with our results.  
The M-T relation obtained from simulations with
 and without pre-heating is similar. However, in pre-heating models, the 
$\MgT$ relation becomes steeper than the $M-T$ relation, due to 
systematic variations of $f_{\rm gas}$ with T.
 However in the work by Bialek \etal (\cite{bialek}) only 7 clusters 
have temperatures above 3.5 keV, with a large
scatter in the $f_{\rm gas}$--T relation (figure 13, their work).  Larger
samples of simulated hot clusters and more detailed studies of their internal
shape are necessary to see if pre-heating can contribute to the bias in
the \MgT relation via $f_{\rm gas}$ variation, without introducing
structural variations, which is inconsistent with the observations.

\section{Conclusion}

Studying the emission measure profiles of a sample of 
clusters ($\kT > 3.5 \keV$)
we found a strong indication that the \MgT relation of 
hot clusters deviates from theoretical expectations. Our 
results suggest $\Mg \propto T^{1.94}$. This relationship together with
the fact that clusters show self-similar profiles can explain the
 observed \LxT relation, which also differs from theoretical expectations. 

This result is in agreement with previous work such as
Mohr \etal (\cite{mohr2}) or Vikhlinin \etal (\cite{vikhlinin}), based on the 
direct measurement of $M_{gas}$ in clusters.  However, our study is more
sensitive to deviations from the predicted \MgT relation since the
emission measure is, with $EM\propto M_{gas}^2$, strongly coupled to
the gas mass content.  Observations from XMM-Newton or Chandra of the
temperature structure are required to assess whether the total mass
also diverges from the predicted scalings.

\begin{acknowledgements}
We would like to thank the referee, Brad Holden, for his very useful comments
to improve the paper.
\end{acknowledgements}


\begin{thebibliography}{}

\bibitem[1989]{abell} Abell, G.O., Corwin, H.G. Jr., Olowin, R.P. 1989, ApJS, 
70, 1

\bibitem[1998]{allen} Allen, S.W., Fabian, A.C. 1998, MNRAS, 297, L57

\bibitem[1999]{arnaud} Arnaud, M., Evrard, A.E. 1999, MNRAS, 305, 631

\bibitem[2001]{bialek} Bialek, J.J., Evrard, A.E., Mohr, J.J. 2001, ApJ
submitted, astro-ph/0010584

\bibitem[1998]{bryan} Bryan, G.L., Norman, M.L. 1998, ApJ, 495, 80

\bibitem[1999]{colafrancesco} Colafrancesco, S. 1999,
in {\it Diffuse Thermal and Relativistic Plasma
in Galaxy Clusters}, eds. H. B\"ohringer, L. Feretti, Sch\"ucker P. MPE Report
271

\bibitem[1998]{eke} Eke, V.R., Navarro, J.F., Frenk, C.S. 1998, ApJ, 503, 569

\bibitem[1995]{elbaz}Elbaz, D., Arnaud, M., B\"ohringer, H. 1995, A\&A, 293, 
337

\bibitem[2000]{ettori}  Ettori, S., Bardelli, S., De Grandi, S., Molendi, S., 
Zamorani, G., Zucca, E. 2000, MNRAS, 318, 239.     

\bibitem[1991]{evrard1} Evrard, A.E., Henry, J.P 1991, ApJ, 383, 95

\bibitem[1996]{evrard2} Evrard, A.E., Metzler, C.A., Navarro, J.F. 1996, ApJ, 
469, 494

\bibitem[1986]{kaiser1} Kaiser, N. 1986, MNRAS, 222, 323

\bibitem[1991]{kaiser2} Kaiser, N. 1991, ApJ, 383, 104

\bibitem[2000]{lloyd} Lloyd-Davies, E.J., Ponman, T.J., Cannon, D.B. 2000, 
MNRAS, 315, 689

\bibitem[2000]{loewenstein} Loewenstein, M. 2000, ApJ, 532, 17

\bibitem[1998]{markevitch1} Markevitch, M., Forman, W.R., Sarazin, C.L., 
Vikhlinin, A. 1998, ApJ 503, 77

\bibitem[1998]{markevitch2} Markevitch, M. 1998, ApJ 504, 27

\bibitem[2001]{mathiesen} Mathiesen, B.F., Evrard, A.E. 2001, ApJ, 546, 100

\bibitem[1997]{mohr1} Mohr, J.J., Evrard, A.E. 1997, ApJ, 491, 38

\bibitem[1999]{mohr2} Mohr, J.J., Mathiesen, B., Evrard, A.E. 1999, 517, 627

\bibitem[2000]{mohr3} Mohr, J.J., Reese, E.D., Ellingson, E., Lewis, A.D., 
Evrard, A.E. 2000, ApJ, 544, 109

\bibitem[1997]{navarro} Navarro, J.F., Frenk, C.S., White, S.D.M. 1997, ApJ, 
490, 493

\bibitem[1999]{neumann} Neumann, D.M., Arnaud, M. 1999, A\&A, 348, 711 (paper 
I)

\bibitem[1999]{ponman} Ponman, T.J., Cannon, D.B., Navarro, J.F. 1999, Nature 
397, 135

\bibitem[1993]{press} Press, W.H., Teukolsky, S.A., Vetterling, W.T., Flannery,
B.P. 1993, Numerical Recipes in C : The Art of Scientific Computing,
Cambridge University Press

\bibitem[1997]{teyssier} Teyssier, R., Chi\`eze, J.P., Alimi, J.M. 1997, ApJ,
 480, 36

\bibitem[2001]{tozzi} Tozzi, P., Norman, C.  2001, ApJ, 546, 63

\bibitem[1999]{vikhlinin} Vikhlinin, A., Forman, W., Jones, C. 1999, ApJ, 
525, 47







\end{thebibliography}
\end{document}